\documentstyle[aps,eqsecnum,floats]{revtex}
\newcommand{\speq}{\stackrel{*}{=}}
\newcounter{prblmcnt}
\newenvironment{problem}
  {\stepcounter{prblmcnt} \vspace*{\baselineskip} \hrule 
  \vspace*{\baselineskip} {\bf Problem \arabic{prblmcnt}.} }
  {\vspace*{\baselineskip} \hrule \vspace*{\baselineskip}}
\begin{document}
\draft
\twocolumn[\hsize\textwidth\columnwidth\hsize\csname 
           @twocolumnfalse\endcsname
\title{An introduction to the Lorentz-Dirac equation}  
\author{Eric Poisson}
\address{Department of Physics, University of Guelph, Guelph,
         Ontario, Canada N1G 2W1}
\maketitle
\begin{abstract}
These notes provide two derivations of the Lorentz-Dirac equation. The
first is patterned after Landau and Lifshitz and is based on the
observation that the half-retarded minus half-advanced potential is
entirely responsible for the radiation-reaction force. The second is
patterned after Dirac, and is based upon considerations of
energy-momentum conservation; it relies exclusively on the retarded
potential. The notes conclude with a discussion of the difficulties
associated with the interpretation of the Lorentz-Dirac equation as an 
equation of motion for a point charge. The presentation is essentially
self-contained, but the reader is assumed to possess some elements of
differential geometry (necessary for the second derivation only).    
\end{abstract}
\vskip 2pc]

\narrowtext

\section{Introduction} 

The Lorentz-Dirac equation is an equation of motion for a charged
particle under the influence of an external force as well as its own
electromagnetic field. The particle's world line is described by
the relations $z^\alpha(\tau)$, which give the particle's coordinates
as functions of proper time. We let $u^\alpha(\tau) = dz^\alpha/d\tau$
be the four-velocity, and $a^\alpha(\tau) = du^\alpha/d\tau$ is the
four-acceleration. The Lorentz-Dirac equation is given by 
\begin{equation}
ma^\alpha = F^\alpha_{\rm ext} + \frac{2}{3}\, q^2 
\bigl( \delta^\alpha_{\ \beta} + u^\alpha u_\beta \bigr) 
\dot{a}^\beta,  
\label{1.1}
\end{equation} 
where $m$ is the particle's mass, $q$ its charge, $F^\alpha_{\rm ext}$
the external force, and $\dot{a}^\alpha = da^\alpha/d\tau$. Throughout
the paper we work in relativistic units, so that the speed of light is
equal to unity. In a Lorentz frame in which the particle is
momentarily at rest at the time $\tau$, Eq.~(\ref{1.1}) reduces to   
\begin{equation}
m \bbox{a} = \bbox{F}_{\rm ext} + \frac{2}{3}\, q^2 \dot{\bbox{a}}, 
\label{1.2}
\end{equation}
where $\dot{\bbox{a}} = d\bbox{a}/dt$; bold-faced symbols denote the
spatial components of the corresponding four-vectors. The special
Lorentz frame in which Eq.~(\ref{1.2}) is valid will be referred to as
the ``momentarily comoving Lorentz frame'', or MCLF. Equation
(\ref{1.2}) also gives the nonrelativistic limit of the Lorentz-Dirac
equation.       

My main objective with these notes is to provide a self-contained
derivation of the Lorentz-Dirac equation. In fact, I will present two
such derivations, each of which relying on a different set of
heuristic assumptions. The first derivation is patterned after the
presentation of Landau and Lifshitz \cite{LL}, which I extend
further. The second derivation is patterned after Dirac's classic
paper \cite{Dirac}, although it differs from it in its technical
aspects --- my own presentation requires much less calculational
labour. The level of rigour achieved in these notes matches that of
those two sources. It does not, however, match the level achieved in
what I consider to be the best reference on this topic, the 1980
review article by Teitelboim, Villarroel, and van Weert \cite{TVW}. 
Those three references were my main sources of inspiration. I have
also relied on the books by Jackson \cite{Jackson} and 
Rohrlich \cite{Rohrlich}.     

I begin with a review of the fundamental equations of electromagnetism
in Sec.~II. In Sec.~III I present the Landau-Lifshitz derivation of
the nonrelativistic Lorentz-Dirac equation, Eq.~(1.2). This reveals
that the radiation-reaction part of the vector potential is given by  
$\frac{1}{2} (A^\alpha_{\rm ret} - A^\alpha_{\rm adv})$,
where $A^\alpha_{\rm ret}$ is the retarded solution to the wave
equation satisfied by the vector potential, while $A^\alpha_{\rm adv}$
is the advanced solution. The remaining part of the (retarded)
potential, $\frac{1}{2} (A^\alpha_{\rm ret} + A^\alpha_{\rm adv})$,
does not affect the motion of the charged particle. 

Solving the wave equation for $A_{\rm ret}^\alpha(x)$ is handled via a
Green's function whose entire support is on the past light cone of the
field point $x$. This introduces a mapping between $x$ and a specific
point $z$ on the particle's world line, the point at which it
intersects $x$'s past light cone. The mathematical aspects of this 
mapping are developed in Sec.~IV, in preparation for Sec.~V, in which 
the potential and field of a point charge are calculated. The
Landau-Lifshitz calculation of the radiation-reaction force 
is resumed in Sec.~VI, this time in a fully relativistic context. The 
calculation is once again based on the half-retarded minus 
half-advanced potential, and it produces Eq.~(\ref{1.1}).   

The potential and field at $x$ are naturally expressed in terms of
$u$, the particle's proper time when it encounters $x$'s past light
cone, and $r$, the distance between $x$ and the particle, as measured
at that time in the MCLF. This motivates the introduction of a
(noninertial) coordinate system for flat spacetime, based on $u$, $r$,
and two polar angles. This new system is centered on the accelerated
world line, and Sec.~VII is devoted to its explicit construction. This
prepares the way for Dirac's derivation of Eq.~(\ref{1.1}), which is
presented in Sec.~VIII. Dirac's derivation is based upon
considerations of energy-momentum conservation, and it involves only
the retarded potential.  

The fact that the Lorentz-Dirac equation involves the derivative of
the acceleration vector introduces considerable difficulties in its 
interpretation as an equation of motion. These difficulties are
reviewed in Sec.~IX; the point of view developed there comes largely
from the excellent discussion contained in Flanagan and Wald
\cite{FW}. These difficulties stem from the basic observation that
point particles cannot be given a fully consistent treatment in a
classical theory of electromagnetism. As a way of resolving these
difficulties, Landau and Lifshitz take the Lorentz-Dirac equation
through a reduction-of-order procedure, in which $q^2/m$ is formally 
treated as a small quantity. This procedure is motivated and described
in detail in Sec.~IX. It produces a modified Lorentz-Dirac equation,  
\begin{equation}     
m a^\alpha = F^\alpha_{\rm ext} + \frac{2}{3} \frac{q^2}{m}\,   
\bigl( \delta^\alpha_{\ \beta} + u^\alpha u_\beta \bigr) 
F^\beta_{{\rm ext}\, ,\gamma} u^\gamma, 
\label{1.3}
\end{equation}
which formally is equivalent to Eq.~(\ref{1.1}), but is free of 
difficulties. In the MCLF, or in the nonrelativistic limit,
Eq.~(\ref{1.3}) reduces to 
\begin{equation}
m \bbox{a} = \bbox{F}_{\rm ext} + \frac{2}{3} \frac{q^2}{m}\, 
\dot{\bbox{F}}_{\rm ext}, 
\label{1.4}
\end{equation}
where $\dot{\bbox{F}}_{\rm ext}$ is the complete (convective) time
derivative of the external force. 

\section{The fundamental equations}

We consider an electromagnetic field $F_{\alpha\beta}$ produced by a
point charge $q$ moving in flat spacetime on a world line
$z^\alpha(\tau)$, where $\tau$ is proper time. The
corresponding current density $j^\alpha$ is given by 
\begin{equation}
j^\alpha(x) = q \int d\tau\, u^\alpha\, \delta(x-z),  
\label{2.1}
\end{equation}
where $u^\alpha(\tau) = dz^\alpha/d\tau$ is the particle's
four-velocity; the integration is over the complete world line, and
the $\delta$-function is four dimensional. Initially we take the world
line to be arbitrary; our main goal is to find the particle's
equations of motion. 

The Maxwell field equations are 
\begin{equation}
F^{\alpha\beta}_{\ \ \, ,\beta} = 4\pi j^\alpha.
\label{2.2}
\end{equation}
If we express the electromagnetic field in terms of a vector
potential,
\begin{equation}
F_{\alpha\beta} = A_{\beta,\alpha} - A_{\alpha,\beta},
\label{2.3}
\end{equation}
and if we adopt the Lorentz gauge, $A^\alpha_{\ ,\alpha} = 0$, then
the field equations take the simple form 
\begin{equation}
\Box A^\alpha = -4\pi j^\alpha,
\label{2.4}
\end{equation}
in which $\Box = \eta^{\alpha\beta} \partial_\alpha \partial_\beta$ is 
the wave operator. The Minkowski metric $\mbox{diag}(-1,1,1,1)$ is
denoted $\eta_{\alpha\beta}$, and $\eta^{\alpha\beta}$ is its inverse.  

This equation can be solved with the help of a Green's function
$G(x,x')$, which satisfies
\begin{equation}
\Box G(x,x') = -4\pi \delta(x-x');
\label{2.5}
\end{equation}
the solution is $A^\alpha(x) = \int G(x,x') j^\alpha(x')\, d^4 x' +
A^\alpha_{\rm hom}(x)$, in which the second term represents a
solution to the homogeneous equation. The {\it retarded} Green's
function can be expressed as 
\begin{eqnarray}
G_{\rm ret}(x,x') &=& \theta(t-t')\, \delta(\sigma) 
\label{2.6} \\
&=& \frac{\delta(t - t' - |\bbox{x} - \bbox{x'}|)}
{|\bbox{x} - \bbox{x'}|}.  
\label{2.7}
\end{eqnarray}
In the first expression, the quantity $\sigma$ is defined by   
\begin{equation}
\sigma(x,x') = \frac{1}{2}\, \eta_{\alpha\beta} (x-x')^\alpha
(x-x')^\beta;
\label{2.8}
\end{equation}
it is half the spacetime interval between the points $x$ and
$x'$. The retarded Green's function has support only on the past light
cone of the field point $x$. The second expression is obtained from
the first by factorizing $\sigma$; $\bbox{x}$ is the spatial
projection of the four-vector $x^\alpha$ and $|\bbox{x}|$ is its
magnitude. The {\it advanced} Green's function is given by  
\begin{eqnarray}
G_{\rm adv}(x,x') &=& \theta(t'-t) \delta(\sigma) 
\label{2.9} \\
&=& \frac{\delta(t - t' + |\bbox{x} - \bbox{x'}|)}
{|\bbox{x} - \bbox{x'}|}.  
\label{2.10}
\end{eqnarray}
It has support on the future light cone of the field point $x$. 

\begin{problem}
Show that Eqs.~(\ref{2.6})--(\ref{2.10}) give the correct expressions
for the retarded and advanced Green's functions. 
\end{problem}

The field equations (\ref{2.2}) imply the charge-conservation equation 
$j^\alpha_{\ ,\alpha} = 0$. It is easy to check that this equation is
satisfied by the current density of Eq.~(\ref{2.1}); the calculation
requires the identity
\begin{equation}
u^\alpha \partial_\alpha \delta(x-z) = -\frac{d}{d\tau}\,
\delta(x-z), 
\label{2.11}
\end{equation}
which is most easily established in the particle's momentarily
comoving Lorentz frame (MCLF) --- the frame in which the particle is 
momentarily at rest at the time $\tau$. Another relevant conservation
equation is
\begin{equation}
(T_{\rm em}^{\alpha\beta} + T_{\rm part}^{\alpha\beta})_{,\beta} = 0, 
\label{2.12}
\end{equation}
where
\begin{equation}
T_{\rm em}^{\alpha\beta} = \frac{1}{4\pi}\, \biggl(
F^{\alpha\mu} F^\beta_{\ \mu} - \frac{1}{4}\, g^{\alpha\beta}
F^{\mu\nu} F_{\mu \nu} \biggr)
\label{2.13}
\end{equation}
is the electromagnetic field's stress-energy tensor, while
\begin{equation}
T_{\rm part}^{\alpha\beta} = m 
\int d\tau\, u^\alpha u^\beta\, \delta(x-z) 
\label{2.14}
\end{equation}
is the stress-energy tensor of a point particle with mass $m$. 

It is easy to show that 
\begin{equation}
\partial_\beta T^{\alpha\beta}_{\rm em} 
= - F^\alpha_{\ \beta}\, j^\beta 
= -q \int d\tau\, F^\alpha_{\ \beta} u^\beta\, \delta(x-z),
\label{2.15}
\end{equation}
while
\begin{equation}
\partial_\beta T_{\rm part}^{\alpha\beta} 
= m\int d\tau\, a^\alpha\, \delta(x-z),
\label{2.16}
\end{equation}
where $a^\alpha(\tau) = du^\alpha/d\tau$ is the particle's
acceleration. Substituting Eq.~(\ref{2.15}) and (\ref{2.16}) into
Eq.~(\ref{2.12}) yields the Lorentz-force equation, 
\begin{equation}
m a^\alpha = q F^\alpha_{\ \beta} u^\beta.
\label{2.17}
\end{equation}
In this equation, $F_{\alpha\beta}$ must be evaluated at the point
$x = z(\tau)$. Since $F_{\alpha\beta}$ is obviously singular on the
world line, Eq.~(\ref{2.17}) is not very meaningful as it stands. Our
aim is therefore to make sense of this equation.  

\begin{problem}
Show that Eqs.~(\ref{2.2}) and (\ref{2.17}) follow from the following
action principle:  
\[
S = \int \biggl( -\frac{1}{16\pi}\, F_{\alpha\beta} F^{\alpha\beta} 
+ A_\alpha j^\alpha \biggr)\, d^4 x - m \int d\tau. 
\]
Here, $j^\alpha = q\int d\lambda\, \dot{z}^\alpha\,
\delta(x-z)$ and $d\tau = (-g_{\alpha\beta} \dot{z}^\alpha
\dot{z}^\beta)^{1/2}\, d\lambda$, where $\dot{z}^\alpha =
dz^\alpha/d\lambda$; $\lambda$ is an arbitrary parameter on the  
world line.  
\end{problem} 

In a specific Lorentz frame, $A^\alpha$ can be decomposed into a
scalar potential $\Phi$ and a vector potential $\bbox{A}$. Similarly, 
$F_{\alpha\beta}$ can be decomposed into an electric field $\bbox{E}$
and a magnetic field $\bbox{B}$. In this Lorentz frame,
Eq.~(\ref{2.3}) becomes
\begin{equation}
\bbox{E} = -\bbox{\nabla} \Phi - \frac{\partial \bbox{A}}{\partial t},
\qquad 
\bbox{B} = \bbox{\nabla} \times \bbox{B},
\label{2.18}
\end{equation}
and Eq.~(\ref{2.17}) reduces to 
\begin{equation}
m \bbox{a} = q ( \bbox{E} + \bbox{v} \times \bbox{B} ),
\label{2.19}
\end{equation}
where $\bbox{v} = d\bbox{z}/dt$ and $\bbox{a} = d\bbox{v}/dt$. Also,
$j^\alpha$ can be decomposed into a charge density $\rho$ and a
current density $\bbox{j}$; these are obtained by changing the
variable of integration in Eq.~(\ref{2.1}) to $z^0$:     
\begin{equation}
\rho(t,\bbox{x}) = q\, \delta(\bbox{x} - \bbox{z}), \qquad
\bbox{j}(t,\bbox{x}) = q \bbox{v}\, \delta(\bbox{x} - \bbox{z}). 
\label{2.20}
\end{equation}
In these expressions, $\bbox{z}$ and $\bbox{v}$ are considered to be
functions of $t$. 

\section{Radiation reaction for slowly moving charges}   

The assumption that the charge is moving slowly gives rise to a clear   
identification of the radiation-reaction part of the electromagnetic
field $F_{\alpha\beta}$. In this section we will show that the part of
the vector potential which is responsible for the radiation reaction
is 
\begin{equation}
A^\alpha_{\rm rr}(x) = \frac{1}{2}\, \bigl[ A^\alpha_{\rm ret}(x) 
- A^\alpha_{\rm adv}(x) \bigr]. 
\label{3.1}
\end{equation}
The remaining part, $\frac{1}{2}( A^\alpha_{\rm ret} 
+ A^\alpha_{\rm adv} )$, must be associated with the particle's
Coulomb field and does not influence the particle's motion. The two
parts add up to $A^\alpha_{\rm ret}$, the correct (retarded) solution
to the Maxwell field equations. The radiation-reaction potential
(\ref{3.1}) gives rise to a radiation-reaction field
$F^{\alpha\beta}_{\rm rr}$, and it is this quantity that must be
substituted to the right-hand side of Eq.~(\ref{2.17}) to obtain the
correct equations of motion. Our considerations in this section will
be limited to the nonrelativistic limit; the general case will be
considered in the following sections. 

Using Eqs.~(\ref{2.7}) and (\ref{2.10}), we find that the retarded and
advanced solutions to Eq.~(\ref{2.4}) are 
\begin{equation}
A^\alpha_\epsilon(t,\bbox{x}) = \int \frac{j^\alpha(t - \epsilon 
|\bbox{x} - \bbox{x'}|, \bbox{x'})}{|\bbox{x} - \bbox{x'}|}\, d^3 x',  
\label{3.2}
\end{equation}
up to the possible addition of a solution to the homogeneous equation;
such a term would give rise to an external field 
$F^{\alpha\beta}_{\rm ext}$ which can always be introduced at a later
stage. The parameter $\epsilon$ is equal to $+1$ for the retarded
solution, and to $-1$ for the advanced solution. For the time being we
leave $j^\alpha$ arbitrary; it could describe an extended charge
distribution. We assume that this charge distribution moves slowly:
Let $r_c$ be a characteristic length scale and $t_c$ a characteristic
time scale associated with the source's motion; we assume 
$v \equiv r_c/t_c \ll 1$. We will use this condition to simplify the
appearance of Eq.~(\ref{3.2}). 

Suppose first that we want to evaluate $A^\alpha_\epsilon$ far from
the source. Under the condition $ r \equiv |\bbox{x}| \gg
|\bbox{x'}|$, we have $|\bbox{x} - \bbox{x'}| = r - \bbox{n} \cdot
\bbox{x'} + O(r^{-1})$, where $\bbox{n} = \bbox{x}/r$ is a unit vector
pointing in the direction of $\bbox{x}$. If we now expand $j^\alpha$
in a Taylor series about $t-\epsilon r$, treating $\bbox{n} \cdot
\bbox{x'}$ as a small quantity, we obtain
\[
j^\alpha(t - \epsilon |\bbox{x} - \bbox{x'}|, \bbox{x'}) = 
\sum_{l=0}^\infty \frac{\epsilon^l}{l!}\, 
(\bbox{n} \cdot \bbox{x'})^l\, 
\frac{\partial^l}{\partial w^l}\, 
j^\alpha(w, \bbox{x'}), 
\]
where $w = t - \epsilon r$ represents retarded time if $\epsilon = 1$
and advanced time if $\epsilon = -1$. The slow-motion approximation
ensures than in this sum, the contribution at order $l$ is a factor 
$(r_c/t_c)^l = v^l$ smaller than the leading term
($l=0$). Substituting this expression for $j^\alpha$ into
Eq.~(\ref{3.2}) yields a multipole expansion for the vector potential: 
\begin{equation}
A^\alpha_\epsilon(t,\bbox{x}) = \frac{1}{r}\,    
\sum_{l=0}^\infty \frac{\epsilon^l}{l!}\,
n_L \frac{d^l}{d w^l}\, \int j^\alpha(w, \bbox{x'})\, 
x^{\prime L}\, d^3 x', 
\label{3.3}
\end{equation}
where we use a multi-index $L$ to denote a product of $l$ identical
factors. For example, $n_L = n_{a_1} n_{a_2} \cdots n_{a_l}$, and
summation over a repeated multi-index is understood. If
$\epsilon = 1$, then $r A^\alpha$ is a function of $t - r$ and the
vector potential represents an {\it outgoing} wave; to such a wave 
corresponds an outward flux of energy, and outgoing waves therefore
{\it remove} energy from the source. On the other hand, if $\epsilon =
-1$ then $r A^\alpha$ is a function of $t + r$ and the vector
potential represents an {\it ingoing} wave; to such a wave corresponds
an inward flux of energy, and ingoing waves therefore {\it provide}
energy to the source. This distinction between the retarded ($\epsilon
= 1$) and advanced ($\epsilon = -1$) solutions is very important for
radiation reaction. 

\begin{problem}
First, show that to leading order in a slow-motion approximation,  
Eq.~(\ref{3.3}) reduces to 
\[
r \Phi_\epsilon(t,\bbox{x}) = q + \epsilon\, \bbox{n} \cdot
\dot{\bbox{p}}, \qquad
r \bbox{A}_\epsilon(t,\bbox{x}) = \dot{\bbox{p}},
\]
where $q = \int \rho\, d^3 x$ is the total charge and $\bbox{p} = 
\int \rho\, \bbox{x}\, d^3 x$ the dipole moment of the charge
distribution. The total charge is a constant, and $\bbox{p}$ is a
function of $w = t -\epsilon r$; an overdot denotes differentiation
with respect to $w$. Second, show that the magnetic field is given by
\[
r \bbox{B}_\epsilon = -\epsilon\, \bbox{n} \times \ddot{\bbox{p}},
\]
and that the electric field satisfies $\bbox{E} = \epsilon\, \bbox{B}
\times \bbox{n}$. Third, calculate the Poynting vector and show that
the rate at which energy is flowing out of a sphere of radius $r$ is
given by  
\[
\frac{dE}{dw} = \epsilon\, \frac{2}{3}\, \ddot{\bbox{p}}^2.
\]
Thus, we have an outward flux if $\epsilon = 1$ and an inward flux if
$\epsilon = -1$. For a point charge, $\bbox{p} = q \bbox{z}$ and
$dE/dw = \epsilon\, \frac{2}{3} q^2 \bbox{a}^2$. This is the
well-known Larmor formula. 
\end{problem}

Suppose now that we want to evaluate $A^\alpha_\epsilon$ inside the 
source. In such a situation we may expand $j^\alpha$ in a Taylor
series about $t$, treating $|\bbox{x} - \bbox{x'}|$ as a small
quantity. This gives
\[
j^\alpha(t - \epsilon |\bbox{x} - \bbox{x'}|, \bbox{x'}) = 
\sum_{l=0}^\infty \frac{(-\epsilon)^l}{l!}\, 
|\bbox{x} - \bbox{x'}|^l\, 
\frac{\partial^l}{\partial t^l}\, 
j^\alpha(t, \bbox{x'}), 
\]
and again we see that the contribution at order $l$ is a factor $v^l$
smaller than the leading term. In this expression we notice that
$(-\epsilon)^l = +1$ if $l$ is even, while $(-\epsilon)^l = -\epsilon$
if $l$ is odd. Substituting this into Eq.~(\ref{3.2}) gives
\begin{eqnarray*}
A^{\alpha}_\epsilon(t,\bbox{x}) &=&  
\sum_{l\ {\rm even}} \frac{1}{l!}\, \frac{\partial^l}{\partial t^l}\, 
\int j^\alpha(t,\bbox{x'})\, |\bbox{x}-\bbox{x'}|^{l-1}\, d^3 x' 
\nonumber \\
& & \mbox{} - \epsilon 
\sum_{l\ {\rm odd}} \frac{1}{l!}\, \frac{\partial^l}{\partial t^l}\, 
\int j^\alpha(t,\bbox{x'})\, |\bbox{x}-\bbox{x'}|^{l-1}\, d^3 x'.
\end{eqnarray*}
In this expression, the first sum gives $\frac{1}{2}( 
A^\alpha_{\rm ret} + A^\alpha_{\rm adv} )$, and this contribution to
$A^\alpha_\epsilon$ is the same irrespective of the nature of the
radiation at infinity, that is, whether the waves are outgoing or
ingoing. This contribution to the vector potential cannot be
responsible for the radiation reaction. On the other hand, the
second sum gives $\frac{1}{2}(A^\alpha_{\rm ret}-A^\alpha_{\rm adv})$,
and this contribution to $A^\alpha_\epsilon$ changes sign under a
change of boundary conditions at infinity. It is clearly this
contribution to the vector potential that is responsible for the
radiation reaction. 

We therefore conclude that the radiation-reaction potential is
indeed given by Eq.~(\ref{3.1}). In a slow-motion approximation, this
reduces to
\begin{equation}
A^{\alpha}_{\rm rr}(t,\bbox{x}) = - 
\sum_{l\ {\rm odd}} \frac{1}{l!}\, \frac{\partial^l}{\partial t^l}\, 
\int j^\alpha(t,\bbox{x'})\, |\bbox{x}-\bbox{x'}|^{l-1}\, d^3 x'. 
\label{3.4}
\end{equation}
The leading contribution to the scalar potential comes from $l=3$ --- 
the $l=1$ term vanishes identically by virtue of charge
conservation. Thus, 
\[
\Phi_{\rm rr}(t,\bbox{x}) = -\frac{1}{3!}\, 
\frac{\partial^3}{\partial t^3} \int \rho(t,\bbox{x'})\,  
|\bbox{x}-\bbox{x'}|^{2}\, d^3 x'.
\]
The leading contribution to the vector potential comes from $l=1$:  
\[
\bbox{A}_{\rm rr}(t,\bbox{x}) = - \frac{\partial}{\partial t} 
\int \bbox{j}(t,\bbox{x'})\, d^3 x'.  
\]
Both expressions neglect terms which are smaller by factors of order
$v^2$. 

We now specialize to the case of a point charge, and use
Eqs.~(\ref{2.20}). We obtain
\[
\Phi_{\rm rr}(t,\bbox{x}) = -\frac{q}{3!}\, 
\frac{d^3}{d t^3}\, |\bbox{x}-\bbox{z}(t)|^{2}, \quad
\bbox{A}_{\rm rr}(t,\bbox{x}) = - q \frac{d}{dt}\, \bbox{v}(t). 
\]
Evaluating the time derivatives gives
\[
\Phi_{\rm rr}(t,\bbox{x}) = \frac{q}{3}\, (\bbox{x} - \bbox{z})
\cdot \dot{\bbox{a}} - q \bbox{v} \cdot \bbox{a}, \quad
\bbox{A}_{\rm rr}(t,\bbox{x}) = - q \bbox{a},
\]
where $\dot{\bbox{a}} = d\bbox{a}/dt$. Substituting these expressions
into Eqs.~(\ref{2.18}), we obtain the radiation-reaction fields, 
\begin{equation}
\bbox{E}_{\rm rr} = \frac{2}{3}\, q \dot{\bbox{a}}, \qquad
\bbox{B}_{\rm rr} = 0.
\label{3.5}
\end{equation}
Finally, Eq.~(\ref{2.19}) gives us the radiation-reaction force: 
\begin{equation}
\bbox{F}_{\rm rr} = \frac{2}{3}\, q^2 \dot{\bbox{a}}. 
\label{3.6}
\end{equation}
The equation of motion of a charged particle
is therefore $m \bbox{a} = \bbox{F}_{\rm ext} + \frac{2}{3} q^2
\dot{\bbox{a}}$, in which the first term represents an externally
applied force.  

We may now substantiate our previous claim that the radiation-reaction
force is associated with a loss of energy in the source. The rate at
which the force is doing work is given by 
\[
\dot{W} = \bbox{F}_{\rm rr} \cdot \bbox{v} 
= \frac{2}{3}\, q^2 \dot{\bbox{a}} \cdot \bbox{v} 
= \frac{2}{3}\, q^2 \biggl[ \frac{d}{dt} (\bbox{a} \cdot \bbox{v})  
- \bbox{a}^2 \Biggr]. 
\]
Upon averaging, assuming either that the motion is periodic or
unaccelerated at early and late times, we obtain
\begin{equation}
\langle \dot{W} \rangle = - \frac{2}{3}\, q^2 \bbox{a}^2. 
\label{3.7}
\end{equation}
The quantity appearing on the right-hand side is (minus) the energy
radiated per unit time by the charged particle. We therefore have
energy balance, on the average.     

Equation (\ref{3.6}) is the leading term in an expansion of the
radiation-reaction force in powers of the particle's velocity. A
careful examination of the higher-order terms in Eq.~(\ref{3.4})
reveals that in fact, Eq.~(\ref{3.6}) is {\it exact} in the
particle's MCLF \cite{LL}. To generalize to other Lorentz frames, we
must find the unique tensorial equation that reduces to
Eq.~(\ref{3.6}) in the MCLF. It is easy to check that the correct
answer is 
\begin{equation}
F^\alpha_{\rm rr} = \frac{2}{3}\, q^2 \bigl( \delta^\alpha_{\ \beta} +
u^\alpha u_\beta \bigr) \dot{a}^\beta,
\label{3.9}
\end{equation}
where $\dot{a}^\alpha(\tau) = da^\alpha/d\tau$. The equations of
motion for a point particle of charge $q$ and mass $m$ are therefore  
\[
m a^\alpha = F^\alpha_{\rm ext} + \frac{2}{3}\, q^2 
\bigl( \delta^\alpha_{\ \beta} + u^\alpha u_\beta \bigr)
\dot{a}^\beta, 
\]
where $F^\alpha_{\rm ext}$ is the external force. This is the {\it
Lorentz-Dirac equation}. In the following sections we will derive it
using fully covariant methods.   

\begin{problem}
Show that Eq.~(\ref{3.9}) reduces to Eq.~(\ref{3.7}) in the
momentarily comoving Lorentz frame. Show also that the
radiation-reaction force can be expressed as $F^\alpha_{\rm rr} =
\frac{2}{3} q^2 (\dot{a}^\alpha - a^2 u^\alpha)$, where $a^2 =
a_\alpha a^\alpha$. 
\end{problem}

\section{Light-cone mapping} 

The fully relativistic calculation of the potential and field
associated with a point charge relies heavily on the causal structure
of the retarded Green's function $G_{\rm ret}(x,x')$, which has
support on the past light cone of the field point $x$. The field at
$x$ therefore depends on the state of the charge's motion at the
instant at which its world line intersects the past light cone. Thus,
the past light cone defines a natural mapping between the field-point
$x$ and a specific point $z$ on the world line. In this section we
develop the mathematical tools associated with this mapping. The
results derived here will be used throughout the following sections.  

\begin{figure}
\vspace*{2.5in}
\special{hscale=30 vscale=30 hoffset=30.0 voffset=-30.0
         angle=0.0 psfile=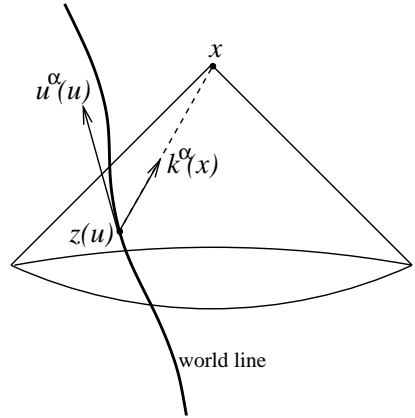}
\caption{Light-cone mapping between an arbitrary field point $x$ and
the point $z(u)$ on the world line.}
\end{figure}

Let $x$ be a field point, and $z(u)$ the point at which the world line
intersects the past light cone of $x$ (see Fig.~1). Here, $u(x)$
denotes the specific value of the particle's proper time $\tau$ which 
corresponds to the intersection point; we shall refer to $u$ as the
{\it retarded time}. Clearly, $u$ is determined by $x$; it is obtained
by solving the equation
\begin{equation}
0 = \sigma(x,u) \equiv \frac{1}{2}\, \eta_{\alpha\beta} 
\bigl[ x^\alpha - z^\alpha(u) \bigr] 
\bigl[ x^\beta - z^\beta(u) \bigr],
\label{4.1}
\end{equation}
which states that the points $x$ and $z(u)$ are linked by a null
geodesic. 

We will need an invariant measure of the distance between $x$ and
$z(u)$. Consider the scalar quantity
\begin{equation}
r(x) = -\eta_{\alpha\beta} \bigl[ x^\alpha - z^\alpha(u) \bigr]\,  
u^\beta(u). 
\label{4.2}
\end{equation}
In the MCLF associated with the particle's motion at the retarded
time, we have $r = t - z^0(u)$, which is just the time required for
light to propagate from $z(u)$ to $x$. Because the speed of light is
set to unity, $r$ is also the spatial distance between these two
points. Thus, the invariant $r(x)$ is the distance between $x$ and
$z(u)$, as measured in the MCLF; we may refer to it as the {\it
retarded distance} between the field point and the particle. Notice
that because $u$ is determined by $x$, there is no need to make the
dependence on $u$ explicit in $r$: we write $r(x)$ and not $r(x,u)$.           

The vector $x^\alpha - z^\alpha(u)$ is a null vector pointing from
$z(u)$ to $x$. It is useful to rescale this vector by a factor
$r^{-1}$, so as to define a new vector, 
\begin{equation}
k^\alpha(x) = \frac{1}{r}\, \bigl[ x^\alpha - z^\alpha(u) \bigr]. 
\label{4.3}
\end{equation}
By virtue of Eqs.~(\ref{4.1}) and (\ref{4.2}), this satisfies 
\begin{equation}
k_\alpha(x) k^\alpha(x) = 0, \qquad 
k_\alpha(x) u^\alpha(u) = -1; 
\label{4.4}
\end{equation}
the second relation provides a convenient normalization for the null
vector.  
 
Because $x$ and $z(u)$ are linked by the light-cone mapping, a change
of field point $x$ generally comes with a change in $u$. (An exception
arises if the displacement is directed along the null geodesic linking
$x$ to $z$.) For infinitesimal displacements, it is easy to relate a
variation in $x$ to the corresponding change in $u$. Suppose that $x$
is displaced to the new field point $x + \delta x$. The new
intersection point in then $z(u + \delta u)$, and these points are
related by Eq.~(\ref{4.1}), $\sigma(x + \delta x, u + \delta u) =
0$. Expanding this to first order in $\delta x$ and $\delta u$, and
using Eqs.~(\ref{4.3}) and (\ref{4.4}), we obtain
$k_\alpha \delta x^\alpha + \delta u = 0$, or
\begin{equation}
\frac{\partial u}{\partial x^\alpha} = - k_\alpha.
\label{4.5}
\end{equation}
This result allows us to efficiently differentiate a function of $x$
which contains an implicit reference to the point $u$. Let $f(x)$
be such a function. We may make the dependence on $u$ explicit by
writing $f(x) = F(x,u)$. Upon differentiation, we momentarily treat
$x$ and $u$ as independent variables, and write $df = 
(\partial F / \partial x^\alpha)\, dx^\alpha + 
(\partial F/ \partial u)\, du$. We then use Eq.~(\ref{4.5}) to obtain
the full derivative with respect to $x^\alpha$: 
\begin{equation}
\frac{\partial f}{\partial x^\alpha} = 
\biggl( \frac{\partial F}{\partial x^\alpha} \biggr)_{\!\! u} -
k_\alpha \biggl( \frac{\partial F}{\partial u} \biggr)_{\!\! x}. 
\label{4.6}
\end{equation}       
This is the differentiation rule under the light-cone mapping.  

As a specific application of Eq.~(\ref{4.6}), it is easy to check that
the derivative of the retarded distance is given by
\begin{equation}
r_\alpha \equiv r_{,\alpha} = - u_\alpha + (1 + r a_k) k_\alpha, 
\label{4.7}
\end{equation}
where $a_k = a_\alpha k^\alpha$ is the component of the acceleration
$a^\alpha = du^\alpha/d\tau$ in the direction of $k^\alpha$. It is
understood that in Eq.~(\ref{4.7}), all world-line quantities (such as
$u^\alpha$ and $a^\alpha$) are to be evaluated at the retarded time
$u$. Notice that Eq.~(\ref{4.7}) implies 
\begin{equation}
k_\alpha(x) r^\alpha(x) = 1.
\label{4.8}
\end{equation}
As another example, we may use Eq.~(\ref{4.6}) to calculate the
derivative of $k^\alpha$. The result is 
\begin{equation}
k_{\alpha,\beta} = \frac{1}{r} \Bigl( g_{\alpha\beta} 
+ k_\alpha u_\beta + u_\alpha k_\beta - k_\alpha k_\beta \Bigr) 
- a_k k_\alpha k_\beta.
\label{4.9}
\end{equation}
This implies that $k^\alpha$ satisfies the geodesic equation,
$k_{\alpha,\beta} k^\beta = 0$. We also note that  
$k^\alpha_{\ ,\alpha} = 2/r$.       

\begin{problem}
Verify Eqs.~(\ref{4.7}) and (\ref{4.9}). Show that $r$ is an affine
parameter on the null geodesic linking $x$ to $z(u)$. 
\end{problem} 

\section{Electromagnetic field of a relativistic point charge}  

The retarded solution to Eq.~(\ref{2.4}), with the current density
given by Eq.~(\ref{2.1}), is $A^\alpha(x) = q \int d\tau\, u^\alpha
G_{\rm ret}(x,z)$, in which both $z^\alpha$ and $u^\alpha$ are
functions of proper time $\tau$; the integration is over the complete
world line. (We again discard the possibility of adding a solution to
the homogeneous equation, which would give rise to an external field
$F^{\alpha\beta}_{\rm ext}$.) Substituting Eq.~(\ref{2.6}), we obtain 
\[
A^\alpha(x) = q \int d\tau\, u^\alpha \theta(t-z^0) \delta(\sigma), 
\]
where $\sigma(x,z)$ is defined by Eq.~(\ref{2.8}). The
$\delta$-function selects two points on the world line, those which
satisfy the relation $\sigma(x,z) = 0$. One solution gives the
retarded time $u$, defined by the condition that $z(u)$ causally
precedes $x$; the other gives the advanced time $v$, defined by the
condition that $x$ precedes $z(v)$. The $\theta$-function rejects the
advanced-time solution, so that $A^\alpha(x)$ depends on the state of
the particle's motion at the retarded time only. 

The integral can be evaluated by changing the variable of integration
to $\sigma$, writing $d\tau = d\sigma/\dot{\sigma}$. We note that
since $\sigma$ goes from negative to positive values as $\tau$ passes 
through the retarded time $u$, $\dot{\sigma}$ is positive at $\tau =
u$. In fact, a quick calculation using Eqs.~(\ref{4.1}) and
(\ref{4.2}) reveals that $\dot{\sigma}(\tau = u) = r(x)$. The vector
potential is therefore given by 
\begin{equation}
A^\alpha(x) = q \frac{u^\alpha(u)}{r(x)}. 
\label{5.1}
\end{equation}
This is the well-known {\it Li\'enard-Wichert} potential. 

The electromagnetic field $F_{\alpha\beta}$ is obtained from the
vector potential by using the differentiation rule of
Eq.~(\ref{4.6}). Using $u_{\alpha,\beta} = - a_\alpha k_\beta$ and
Eq.~(\ref{4.7}), we obtain
\begin{equation}
F_{\alpha\beta} = \frac{2q}{r} \Bigl( a_{[\alpha} k_{\beta]} 
+ a_k\, u_{[\alpha} k_{\beta]} \Bigr) 
+ \frac{2q}{r^2}\, u_{[\alpha} k_{\beta]},
\label{5.2}
\end{equation}
where we have suppressed the dependence of the world-line quantities
(such as $u^\alpha$, $a^\alpha$, and $a_k = a_\alpha k^\alpha$) on the
retarded time $u$. The square brackets denote antisymmetrization of
the indices: $a_{[\alpha} k_{\beta]} = \frac{1}{2} ( a_\alpha k_\beta
- k_\alpha a_\beta )$. (Below, we will use round brackets to denote
symmetrization of the indices.) It is interesting to note that the
part of the electromagnetic field which scales as $r^{-1}$ is
proportional to the particle's acceleration; this is the 
{\it radiative} part of the field. On the other hand, the part 
which scales as $r^{-2}$ does not involve the acceleration; this is
the {\it bound} --- or {\it Coulomb} --- part of the field.         

\begin{problem}
Check the validity of Eq.~(\ref{5.2}). Verify that the electromagnetic
field satisfies the vacuum Maxwell equations, $F^{\alpha\beta}_{\ \
\, ,\beta} = 0$, away from the world line. 
\end{problem}

It is straightforward to substitute Eq.~(\ref{5.2}) into
Eq.~(\ref{2.13}) to calculate the electromagnetic field's
stress-energy tensor. This computation reveals a natural 
decomposition into radiative and bound components, 
\begin{equation}
T_{\rm em}^{\alpha\beta} = T_{\rm rad}^{\alpha\beta}
 + T_{\rm bnd}^{\alpha\beta},
\label{5.3}
\end{equation}
where
\begin{equation}
T_{\rm rad}^{\alpha\beta} = \frac{q^2}{4\pi r^2} \Bigl(
a^2 - {a_k}^2 \Bigr) k^\alpha k^\beta
\label{5.4}
\end{equation}
is the radiative component, and
\begin{eqnarray}
T_{\rm bnd}^{\alpha\beta} &=& \frac{q^2}{2\pi r^3} \biggl[
k^{(\alpha} a^{\beta)} + a_k \Bigl( k^{(\alpha} u^{\beta)} 
- k^\alpha k^\beta \Bigr) \biggr] 
\nonumber \\ & & \mbox{}
+ \frac{q^2}{4\pi r^4} \biggl[ 2k^{(\alpha} u^{\beta)} 
- k^\alpha k^\beta + \frac{1}{2}\, \eta_{\alpha\beta} \biggr]
\label{5.5}
\end{eqnarray}
is the bound --- or Coulomb --- component; we use the notation $a^2 = 
a_\alpha a^\alpha$. This decomposition is meaningful because each
component is separately conserved away from the world line, 
\begin{equation}
\partial_\beta T^{\alpha\beta}_{\rm rad} = 0, \qquad
\partial_\beta T^{\alpha\beta}_{\rm bnd} = 0, \qquad
(r \neq 0).
\label{5.6}
\end{equation}
Furthermore, the interpretation of $T^{\alpha\beta}_{\rm rad}$ as the 
radiative part of the stress-energy tensor is motivated by the fact
that it scales as $r^{-2}$ and is proportional to $k^\alpha k^\beta$.    

\begin{problem}
Check the validity of Eqs.~(\ref{5.3})--(\ref{5.5}). Verify that 
$T^{\alpha\beta}_{\rm rad}$ is separately conserved away from the
world line.
\end{problem}

\section{The radiation-reaction force} 

As we have seen in Sec.~III, the radiation-reaction force can be
calculated on the basis of a radiation-reaction potential equal to
half the difference between the retarded and advanced potentials. The 
retarded potential was calculated in Sec.~V, and given in
Eq.~(\ref{5.1}). A similar calculation reveals that the advanced
potential is
\begin{equation}
A^\alpha_{\rm adv}(x) = q \frac{u^\alpha(v)}{r_{\rm adv}(x)}, 
\label{6.1}
\end{equation}
where $v$ is the advanced time, determined by the relation
$\sigma(x,z) = 0$ and the condition that $x$ precedes $z(v)$, and
\begin{equation}
r_{\rm adv}(x) = -\eta_{\alpha\beta} \bigl[ z^\alpha(v) - x^\alpha
\bigr]\, u^\beta(v)
\label{6.2}
\end{equation}
is the advanced distance. 

To construct the radiation-reaction potential, it is advantageous to
express $A^\alpha_{\rm adv}(x)$ in terms of retarded
quantities. Because we are interested in a field point $x$ very close
to the world line, this transcription is easily carried out by Taylor
expansion. We shall therefore expand $\Delta \tau \equiv v-u$ and
$r_{\rm adv}(x)$ in powers of the retarded distance $r$. It is easy to
see that formally, $\Delta \tau$ and $r$ are of the same order of
smallness. 

We begin by substituting the expansion
\begin{eqnarray*}
z^\alpha(v) &=& z^\alpha + u^\alpha\, \Delta \tau + \frac{1}{2}\,
a^\alpha\, \Delta \tau^2 + \frac{1}{6}\, \dot{a}^\alpha\, 
\Delta \tau^3   
\\ & & \mbox{} 
+ \frac{1}{24}\, \ddot{a}^\alpha\, \Delta \tau^4 + O(\Delta \tau^5),   
\end{eqnarray*}
in which all quantities on the right-hand side are evaluated at the
retarded time $u$, into the relation $\sigma(x,z(v)) = 0$. After some
algebra, using the relations $\sigma(x,z(u)) = 0$ and $x^\alpha - 
z^\alpha(u) = r k^\alpha$, we obtain
\begin{eqnarray*}
0 &=& 2r\, \Delta \tau - (1 + r a_k)\, \Delta \tau^2 
- \frac{1}{3}\, r \dot{a}_k\, \Delta \tau^3 
\\ & & \mbox{}
- \frac{1}{12}\, \bigl( r \ddot{a}_k + a^2 \bigr)\, \Delta \tau^4  
+ O(\Delta \tau^5),
\end{eqnarray*}
where, for example, $\dot{a}_k = \dot{a}_\alpha k^\alpha$. This
relation implies that to leading order, $\Delta \tau = 2 r$ --- the
time delay is twice the spatial distance between the world line and
the field point. Calculating higher-order terms is
straightforward, if slightly tedious. The result is 
\begin{equation}
\Delta \tau = 2 r \Biggl[ 1 - a_k r + 
\biggl( {a_k}^2 - \frac{1}{3}\, a^2 - \frac{2}{3}\, \dot{a}_k \biggr)
r^2 + O(r^3) \Biggr]. 
\label{6.3}
\end{equation}
We recall that $\Delta \tau$ stands for $v-u$, and that all quantities
on the right-hand side refer to the retarded time $u$. 

To express $r_{\rm adv}$ in terms of $r$, we substitute the previous
expansion for $z^\alpha(v)$ and a similar expansion for
$u^\alpha(v)$, 
\[
u^\alpha(v) = u^\alpha + a^\alpha\, \Delta \tau + \frac{1}{2}\,
\dot{a}^\alpha\, \Delta \tau^2 + \frac{1}{6}\, \ddot{a}^\alpha\,
\Delta \tau^3 + O(\Delta \tau^4),  
\]
into Eq.~(\ref{6.2}). (Here it is sufficient to keep terms only up to 
order $\Delta \tau^3$.) After using Eq.~(\ref{6.3}), we obtain
\begin{equation}
r_{\rm adv} = r + \frac{2}{3} \bigl( a^2 + \dot{a}_k \bigr) r^3 +
O(r^4). 
\label{6.4}
\end{equation}
We may also use Eq.~(\ref{6.3}) to express $u^\alpha(v)$ as an
expansion in powers of $r$: 
\begin{equation}
u^\alpha(v) = u^\alpha + 2 a^\alpha\, r + 2 \bigl( \dot{a}^\alpha -
a_k a^\alpha \bigr)\, r^2 + O(r^3).
\label{6.5}
\end{equation}
Again, all quantities on the right-hand side refer to the retarded 
time $u$.  

\begin{problem}
Check the validity of Eqs.~(\ref{6.3})--(\ref{6.5}). Give a physical
interpretation to the fact that $r_{\rm adv} = r$ if the motion is
unaccelerated.
\end{problem}  

The advanced potential is obtained by substituting Eqs.~(\ref{6.4}) 
and (\ref{6.5}) into Eq.~(\ref{6.1}). The result is
\begin{eqnarray*}
A^\alpha_{\rm adv}(x) &=& q \frac{u^\alpha}{r} + 2 q a^\alpha + 2 q 
\biggl[ \dot{a}^\alpha - a_k a^\alpha 
\\ & & \mbox{}
- \frac{1}{3} \Bigl( a^2 + \dot{a}_k \Bigr) u^\alpha \biggr] r  
+ O(r^2). 
\end{eqnarray*} 
The radiation-reaction potential is therefore
\begin{eqnarray}
A_{\rm rr}^\alpha(x) &=& \frac{1}{2}\, \bigl[ A^\alpha_{\rm ret}(x) 
- A^\alpha_{\rm adv}(x) \bigr] \nonumber \\
&=& - q a^\alpha - q \biggl[ \dot{a}^\alpha - a_k
a^\alpha - \frac{1}{3} \Bigl( a^2 + \dot{a}_k \Bigr) u^\alpha \biggr] 
r 
\nonumber \\ & & \mbox{} 
+ O(r^2). 
\label{6.6}
\end{eqnarray} 
The terms of order $r^2$ and higher will not contribute to the
radiation-reaction force, because after differentiation they give
rise to terms which vanish on the world line. 

Differentiation of the radiation-reaction potential proceeds with the
help of Eq.~(\ref{4.6}), using such results as Eq.~(\ref{4.7}),
(\ref{4.9}), and $a_{\alpha,\beta} = - \dot{a}_\alpha k_\beta$. The
calculation is straightforward, although it involves a fair amount of
algebra. Many of the terms drop out after antisymmetrization of the
indices, and we obtain 
\begin{equation}
F_{\alpha\beta}^{\rm rr}(z) = - \frac{2}{3}\, q \bigl( \dot{a}_\alpha
u_\beta - u_\alpha \dot{a}_\beta \bigr).  
\label{6.7}
\end{equation}
Substituting this into the Lorentz-force equation yields
\[
F^\alpha_{\rm rr} = q F^\alpha_{{\rm rr}\, \beta} u^\beta =
\frac{2}{3}\, q^2 \bigl( \delta^\alpha_{\ \beta} + u^\alpha u_\beta 
\bigr) \dot{a}^\beta.  
\]
The equations of motion of a charged particle are therefore
\begin{equation}
m a^\alpha = F^\alpha_{\rm ext} + \frac{2}{3}\, q^2 \bigl
( \delta^\alpha_{\ \beta} + u^\alpha u_\beta \bigr) \dot{a}^\beta, 
\label{6.8}
\end{equation}
where $F^\alpha_{\rm ext}$ is an externally applied force. Thus, the
Lorentz-Dirac equation follows directly from the realization that the
half-retarded minus half-advanced potential is responsible for the
radiation reaction. The role of the remaining part of the retarded
potential will be elucidated in Sec.~VIII.  

\begin{problem}
Check that Eq.~(\ref{6.7}) does indeed follow from the
radiation-reaction potential of Eq.~(\ref{6.6}). Check that the
Lorentz-Dirac equation is compatible with the identity
$\eta_{\alpha\beta} u^\alpha a^\beta = 0$. 
\end{problem} 

\section{Retarded coordinates}

Our purpose in this section is to develop the mathematical tools
required in an alternative derivation of the Lorentz-Dirac equation,
to be presented in the following section. We shall construct a
coordinate system for flat spacetime based on the retarded time $u$
and retarded distance $r$; these coordinates will be centered on
the accelerated world line. A broad class of such coordinate systems
was considered by Newman and Unti \cite{NU}; ours is a specific
example.   

The coordinate system $(u,r,\theta^A)$, where $\theta^A =
(\theta,\phi)$ are two polar angles, is constructed as follows. We
select the point $z(u)$ on the world line, and consider the forward
light cone of this point (see Fig.~2). To all the spacetime events
lying on the light cone we assign the same coordinate $u$. The null
cone is generated by null rays emanating from $z(u)$ and radiating in
all possible directions. We may choose a specific ray by selecting two
angles, $\theta^A$, which give the direction of the ray with respect
to a reference axis. To all the spacetime events lying on this null
ray we assign the same coordinates $\theta^A$. Finally, a specific
event on this null ray can be characterized by its affine-parameter
distance from the apex; this is our fourth coordinate $r$.   

\begin{figure}
\vspace*{2.5in}
\special{hscale=30 vscale=30 hoffset=30.0 voffset=-30.0
         angle=0.0 psfile=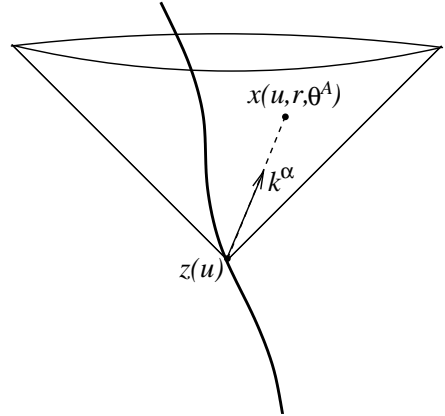}
\caption{A coordinate system centered on an accelerated world line.}  
\end{figure}

The point $x^\alpha(u,r,\theta^A)$ in spacetime is therefore linked 
to the point $z^\alpha(u)$ on the world line by a null geodesic of
affine-parameter length $r$; this null ray belongs to the light cone 
labelled by $u$ and is further characterized by the angles $\theta^A$
specifying its direction on the cone. Let the vector
$k^\alpha(u,\theta^A)$ be tangent to this null ray, which 
then admits the description $dx^\alpha/dr = k^\alpha$. Because
$k^\alpha$ is a null vector, its normalization is arbitrary, and we
are free to impose
\begin{equation}
\eta_{\alpha\beta} k^\alpha(u,\theta^A) u^\beta(u) = -1;
\label{7.1}
\end{equation}
this also determines the normalization of the affine parameter
$r$. Its interpretation as the retarded distance between $x$ and
$z(u)$ comes from the integrated form of the geodesic equation,
\begin{equation}
x^\alpha(u,r,\theta^A) = z^\alpha(u) + r k^\alpha(u,\theta^A),
\label{7.2}
\end{equation}
which is identical to Eq.~(\ref{4.3}). Notice, however, the
differences in point of view: In Sec.~IV, $k^\alpha$ was considered to
be a function of $x$; here, it is a function of $u$ and $\theta^A$ 
which, together with $r$, {\it determine} the point $x$. 

Equation (\ref{7.2}) is the desired transformation between the
inertial system $x^\alpha$ and the (noninertial) retarded coordinates 
$(u,r,\theta^A)$. So far, the construction is not unique: we have yet
to decide how the angles $\theta^A$ are to be defined. In other words,
we must give a prescription to locate the polar axis on each of the
light cones; the angles $\theta^A$ will then be the polar angles
associated with this axis. 

We adopt the following prescription: We consider a Lorentz frame
$(t,x,y,z)$ which is momentarily comoving with the particle at the
time $\tau = u$. In this frame, which we shall call the $u$-frame,
$u^\alpha(u) \speq \delta^\alpha_{\ t}$, where ``$\speq$''
means ``equals in the specified Lorentz frame''. We orient the spatial
axes so that the (purely spatial) acceleration vector points in the
direction of the polar axis, $a^\alpha(u) \speq a(u)\,
\delta^\alpha_{\ z}$, where $a(u)$ is the magnitude of the
acceleration vector. Repeating this construction for every point on
the world line gives us a polar axis on each of the null cones, and
the angles $\theta^A$ will refer to this choice of axis. We let 
\begin{equation}
k^\alpha(u,\theta^A) \speq (1, \sin\theta\, \cos\phi, 
\sin\theta\, \sin\phi, \cos\theta),
\label{7.3}
\end{equation}
which is compatible with the constraints coming from Eq.~(\ref{7.1})
and the fact that $k^\alpha$ is null. 

\begin{problem}
Derive the following relations, which will be required in Sec.~VII:   
\begin{eqnarray*}
\frac{1}{4\pi} \int k^\alpha\, d\Omega &=& u^\alpha, \\
\frac{1}{4\pi} \int k^\alpha k^\beta\, d\Omega &=& \frac{1}{3}\,
g^{\alpha\beta} + \frac{4}{3}\, u^\alpha u^\beta, \\
\frac{1}{4\pi} \int k^\alpha k^\beta k^\gamma\, d\Omega &=&
\frac{1}{3}\, \bigl( u^\alpha g^{\beta\gamma} + u^\beta
g^{\alpha\gamma} + u^\gamma g^{\alpha\beta} \bigr) 
\nonumber \\ & & \mbox{}
+ 2u^\alpha u^\beta u^\gamma,
\end{eqnarray*}
where $d\Omega = \sin\theta\, d\theta d\phi$ is the element of solid
angle. To reduce the work involved, express the null vector as
$k^\alpha = u^\alpha + n^\alpha$, where $n^\alpha$ is a unit vector
orthogonal to $u^\alpha$. Rely on symmetry to argue that $\int
n^\alpha\, d\Omega$ must vanish, and show that $\int n^\alpha
n^\beta\, d\Omega$ must be proportional to 
$g^{\alpha\beta} + u^\alpha u^\beta$.  
\end{problem} 

Equation (\ref{7.3}) gives an explicit expression for
$k^\alpha(u,\theta^A)$ in the $u$-frame. In the Lorentz frame
$(t',x',y',z')$ associated with the time $u'$, defined so that
$u^\alpha(u') \speq \delta^\alpha_{\ t'}$ and $a^\alpha(u') \speq
a(u')\, \delta^\alpha_{\ z'}$, $k^\alpha(u',\theta^A)$ would take
exactly the same form. This fact can be used to obtain a useful
differential equation for $k^\alpha(u,\theta^A)$, which will allow us
to work in a single Lorentz frame. To derive this, we take $u' = u +
\delta u$ and work to first order in $\delta u$. The relative
velocity between the two Lorentz frames is $u^\alpha(u+\delta u) -
u^\alpha(u) = a^\alpha(u)\, \delta u$, which allows us to deduce the
components of $u^\alpha(u+\delta u)$ in the $u$-frame: 
$u^\alpha(u+\delta u) \speq (1,0,0,a\, \delta u)$; we see that the
boost parameter is $v = a(u)\, \delta u$. To first order in $v$, the
Lorentz transformation between the two frames is given by 
$t' = t - v z$, $x' = x$, $y' = y$, and $z' = z - vt$. With this we
can calculate the components of $k^\alpha(u + \delta u)$ in the
$u$-frame: Using Eq.~(\ref{7.3}), we find $k^t(u + \delta u) \speq 1 +
a(u) \cos\theta\, \delta u$, $k^x(u + \delta u) \speq 
\sin\theta\, \cos\phi$, $k^y(u + \delta u) \speq 
\sin\theta\, \sin\phi$, and $k^z(u + \delta u) \speq 
\cos\theta + a(u)\, \delta u$. These results imply 
\[
\frac{\partial k^\alpha}{\partial u} \speq (a\cos\theta,0,0,a) 
\speq (a_k\, u^t, 0, 0, a^z). 
\]
An equivalent tensorial equation is 
\begin{equation}
\frac{\partial k^\alpha}{\partial u} = a_k\, u^\alpha + a^\alpha;
\label{7.4}
\end{equation}
it holds in an arbitrary Lorentz frame. With the initial data of
Eq.~(\ref{7.3}), Eq.~(\ref{7.4}) allows us to find
$k^\alpha(u,\theta)$ at all times.  

\begin{problem}
Derive the relations $k_\alpha \partial k^\alpha/\partial u = 0$ and
$u_\alpha \partial k^\alpha/\partial u + a_k = 0$, and show that
Eq.~(\ref{7.4}) is compatible with them.
\end{problem}

We now turn to the calculation of the metric in the coordinates
$(u,r,\theta^A)$. Differentiation of Eq.~(\ref{7.2}) yields
\begin{equation}
dx^\alpha = \bar{u}^\alpha\, du + k^\alpha\, dr + e^\alpha_A\,
d\theta^A, 
\label{7.5}
\end{equation}
where 
\begin{equation}
\bar{u}^\alpha = u^\alpha + r \frac{\partial k^\alpha}{\partial u} =
(1 + r a_k) u^\alpha + r a^\alpha
\label{7.6}
\end{equation}
and
\begin{equation}
e^\alpha_A = r \frac{\partial k^\alpha}{\partial \theta^A}. 
\label{7.7}
\end{equation}
The metric follows by substituting Eq.~(\ref{7.5}) into $ds^2 =
\eta_{\alpha\beta}\, dx^\alpha dx^\beta$ and calculating all relevant 
inner products between $\bar{u}^\alpha$, $k^\alpha$, and
$e^\alpha_A$. For example, the components of $e^\alpha_A$ in the
$u$-frame can easily be obtained from Eq.~(\ref{7.3}), and we infer 
$k_\alpha e^\alpha_A = 0 = u_\alpha e^\alpha_A$, as well as 
\[
\eta_{\alpha\beta} e^\alpha_A e^\beta_B =
\mbox{diag}(r^2,r^2\sin^2\theta);
\]
these results hold in any Lorentz frame. At the end of a
straightforward computation, we obtain
\begin{eqnarray}
ds^2 &=& -\bigl[ (1 + r a_k)^2 - r^2 a^2 \bigr]\, du^2 - 2\, dudr 
\nonumber \\ & & \mbox{}
+ 2r a_A\, dud\theta^A + r^2\, d\Omega^2,
\label{7.8}
\end{eqnarray}
where $a^2 = a_\alpha a^\alpha$, $a_k = a_\alpha k^\alpha$, $a_A =
a_\alpha e^\alpha_A$, and $d\Omega^2 = d\theta^2 + \sin^2\theta\,
d\phi^2$ is the metric on a unit two-sphere.      

When viewed in the retarded coordinates, the metric of flat spacetime
looks horribly complicated --- there are many nondiagonal terms, and
the metric components depend on the coordinates. What is then the
particular advantage of these coordinates? The answer comes from
Sec.~V, in which we saw that the electromagnetic field of a charged
particle is naturally expressed in terms of $u$ and $r$. This is
true as well of the field's stress-energy tensor, which will be
involved in the calculations of Sec.~VIII. In fact, these calculations
will not involve the metric directly. What we shall need, however, is
an expression for $dV$, the four-dimensional volume element. In spite
of the grim appearance of the metric, this is simple: 
\begin{equation}
dV = \sqrt{-g}\, d^4 x = r^2 du dr d\Omega,
\label{7.9}
\end{equation}
where $d\Omega = \sin\theta\, d\theta d\phi$ is the element of solid
angle. This result is established by computing $g$, the determinant of
the matrix formed by the metric elements; the square root of $-g$ is
the Jacobian of the coordinate transformation
$x^\alpha(u,t,\theta^A)$. 

\begin{problem}
Check the validity of Eqs.~(\ref{7.8}) and (\ref{7.9}). Show that
$\sqrt{-g}$ is the Jacobian of the coordinate transformation. 
\end{problem}

Another result we shall need is
\begin{equation}
d\Sigma_\alpha = r_\alpha\, r^2 du d\Omega, 
\label{7.10}
\end{equation}
which gives the (outward-directed) surface element of a three-cylinder
$r = \mbox{constant}$ in Minkowski spacetime. Here, $r_\alpha =
\partial r / \partial x^\alpha$ is the gradient of $r$ in any
coordinate system $x^\alpha$. In the usual Lorentzian coordinates,
$r_\alpha$ was worked out in Eq.~(\ref{4.7}). In the retarded
coordinates, $r_\alpha = \delta^r_{\ \alpha}$. 

The derivation of Eq.~(\ref{7.10}) proceeds as follows. The vectorial 
surface element on a hypersurface $\Sigma$ is defined by 
$d\Sigma_\alpha = n_\alpha dA$, where $n^\alpha$ is the hypersurface's 
unit normal (pointing in the outward direction) and $dA$ is the
three-dimensional surface element. In our application, $\Sigma$ is a
surface of constant $r$ whose induced metric is obtained by putting
$dr=0$ in Eq.~(\ref{7.8}): 
\[
ds^2_{\Sigma} = -\bigl[ (1 + r a_k)^2 - r^2 a^2 \bigr]\, du^2 + 2r 
a_A\, dud\theta^A + r^2\, d\Omega^2. 
\]
The surface element is then given by $dA = \sqrt{-g_{\Sigma}}\, 
d^3 x$, where $g_\Sigma$ is the determinant of the induced metric.  
Because $\Sigma$ is a surface of constant $r$, the unit normal must be
proportional to the gradient of $r$. We therefore let $n_\alpha =
\lambda r_\alpha$, and determine $\lambda$ by making sure that
$n_\alpha$ is properly normalized. This gives $\lambda^{-2} =
g^{\alpha\beta} r_\alpha r_\beta$. In the retarded coordinates,
$\lambda^{-2} = g^{rr}$, a component of the inverse metric. This is
given by $\mbox{cofactor}(g_{rr})/g$, where the cofactor of a matrix
element is the determinant obtained by eliminating the row and column
to which this element belongs. It is easy to see that
$\mbox{cofactor}(g_{rr}) = g_{\Sigma}$, so that $\lambda =
\sqrt{g/g_{\Sigma}}$. Combining these results, we see that
$d\Sigma_\alpha = n_\alpha\, dA$ reduces to
$d\Sigma_\alpha = r_\alpha \sqrt{g}\, d^3 x$, which is just
Eq.~(\ref{7.10}).           

\begin{problem}
A particle moving with uniform acceleration in the $z$ direction has a
world line described by $z^0 = a^{-1} \sinh(a\tau)$, $z^1 = z^2 = 0$,
and $z^3 = a^{-1} \cosh(a\tau)$, where $a$, the magnitude of the
acceleration vector, is a constant. These components of
$z^\alpha(\tau)$ are given in the ``laboratory frame'', in which the
particle is momentarily at rest at $t = 0$.    

Calculate the components of $k^\alpha(u,\theta^A)$ in the laboratory
frame. Show that the transformation to the retarded coordinates is
given by  
\begin{eqnarray*}
t &=& a^{-1} \sinh\lambda + r (\cosh\lambda 
      + \cos\theta\, \sinh\lambda), \\ 
x &=& r \sin\theta\, \cos\phi, \\
y &=& r \sin\theta\, \sin\phi, \\
z &=& a^{-1} \cosh\lambda + r (\sinh\lambda 
      + \cos\theta\, \cosh\lambda),
\end{eqnarray*}
where $\lambda = a u$. Substitute this into the Minkowski metric,
$ds^2 = -dt^2 + dx^2 + dy^2 + dz^2$, and show that you agree with
Eq.~(\ref{7.8}). 
\end{problem}

\section{Dirac's derivation}

Dirac's derivation of the radiation-reaction force is based upon 
considerations of energy-momentum conservation. It goes as 
follows. We enclose the world line within a thin world tube $\Sigma$
--- a three-cylinder whose shape is {\it a priori} arbitrary (see
Fig.~3) --- and we calculate how much electromagnetic-field momentum
flows across this surface per unit proper time. We then demand
that this change in momentum be balanced by a corresponding change in
the particle's momentum, so that the {\it total} momentum is properly 
conserved. Provided that the particle's momentum is correctly 
identified, this analysis yields the Lorentz-Dirac equation.  

\begin{figure}
\vspace*{2.5in}
\special{hscale=30 vscale=30 hoffset=30.0 voffset=-30.0
         angle=0.0 psfile=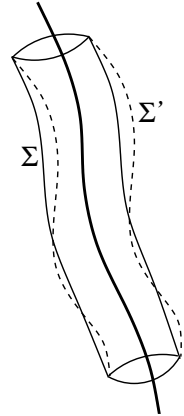}
\caption{The world tube $\Sigma$ enclosing the world line. The dashed
lines show a deformed world tube $\Sigma'$.} 
\end{figure}  

Quite generally, the flux of four-momentum across a hypersurface
$\Sigma$ is given by 
\begin{equation}
\Delta P^\alpha = \int_\Sigma T^{\alpha\beta}\, d\Sigma_\beta,
\label{8.1}
\end{equation}
where $T^{\alpha\beta}$ is any conserved stress-energy tensor and 
$d\Sigma_\alpha$ the outward-directed surface element on $\Sigma$. In
our application, $\Sigma$ will be the world tube of Fig~3. How does
$\Delta P^\alpha$ depend on the shape of the world tube? The answer is
that it does not: the momentum flowing through a deformed tube
$\Sigma'$ is equal to the flux across the original tube $\Sigma$,
provided that the tubes begin and terminate on the same
two-surfaces. (The tubes must therefore have the same ``caps''; we do
not want $\Sigma'$ to be longer than $\Sigma$, thereby allowing more
momentum to go through.) 

The proof of this statement is based on the spacetime formulation of
Gauss' theorem and the fact that the stress-energy tensor is
conserved. Let $\Delta P^{\prime \alpha}$ be the momentum flux across
the deformed tube $\Sigma'$. Let $\cal V$ be the four-dimensional
region between $\Sigma'$ and $\Sigma$; this region's boundary is
denoted $\partial {\cal V}$ and consists of the union of $\Sigma$ and
$\Sigma'$. By Gauss' theorem, we have
\begin{eqnarray*}
\Delta P^{\prime \alpha} - \Delta P^\alpha &=& 
\int_{\partial {\cal V}} T^{\alpha \beta}\, d\Sigma_\beta \\
&=& \int_{\cal V} T^{\alpha\beta}_{\ \ \, ,\beta}\, dV \\
&=& 0,
\end{eqnarray*}  
which proves the assertion. Because the shape of the world tube is
irrelevant, we shall make the simplest choice and take $\Sigma$ to be
a hypersurface of constant $r$. The surface element on this world tube
is given by Eq.~(\ref{7.10}). We will assume that $r$ is small, so
that $\Sigma$ lies in the immediate vicinity of the world line. It
will not be necessary, however, to take the limit $r \to 0$.  

We have seen in Sec.~V that the stress-energy tensor of the
electromagnetic field is naturally decomposed into radiative and bound
parts which are separately conserved away from the world line. By
virtue of Eqs.~(\ref{4.7}) and (\ref{5.4}), the radial component of
the radiative stress-energy tensor is given by 
\[
T^{\alpha\beta}_{\rm rad}\, r_\beta = \frac{q^2}{4\pi r^2} \Bigl( 
a^2 - {a_k}^2 \Bigr) k^\alpha, 
\]
and the flux of radiative momentum is 
\[
\Delta P^\alpha_{\rm rad} = \frac{q^2}{4\pi} \int \bigl( 
a^2 - {a_k}^2 \bigr) k^\alpha\, du d\Omega. 
\]
Notice that the factor $r^{2}$ in the surface element has canceled
out the factor $r^{-2}$ in the radiative stress-energy tensor; the
result is independent of $r$. The rate of momentum change is 
\[
\frac{d P^\alpha_{\rm rad}}{du} = \frac{q^2}{4\pi} \int \bigl( 
a^2 - {a_k}^2 \bigr) k^\alpha\, d\Omega. 
\]
The angular integration can be handled via the results of Problem 10, 
and we arrive at
\begin{equation}
\frac{d P^\alpha_{\rm rad}}{du} = \frac{2}{3}\, q^2 a^2 u^\alpha. 
\label{8.2}
\end{equation}
This is the amount of radiative momentum crossing a surface $r =
\mbox{constant}$ per unit proper time. In the MCLF, this equation
reduces to $dE/dt = \frac{2}{3} q^2 \bbox{a}^2$, the standard Larmor
formula.  

Going back to Eqs.~(\ref{4.7}) and (\ref{5.5}), we find that the
radial component of the bound stress-energy tensor is given by 
\[
T^{\alpha\beta}_{\rm bnd}\, r_\beta = \frac{q^2}{4\pi r^3} 
\biggl[ a^\alpha + a_k \Bigl( u^\alpha - \frac{3}{2} k^\alpha \Bigr)
\biggr] + \frac{q^2}{4\pi r^4} \bigl( u^\alpha - k^\alpha \bigr). 
\]
After integration we notice that the most singular terms, those which
are proportional to $r^{-4}$, go away. The $r^{-3}$ terms, however,
give a nonzero result:
\begin{equation}
\frac{d P^\alpha_{\rm bnd}}{du} = \frac{q^2}{2 r}\, a^\alpha.  
\label{8.3}
\end{equation}
This is the rate of change of the bound momentum. We shall have
to interpret this result.  

\begin{problem}
Check the validity of Eqs.~(\ref{8.2}) and (\ref{8.3}).
\end{problem}

The rate of change of electromagnetic momentum is given by the sum of  
Eqs.~(\ref{8.2}) and (\ref{8.3}), 
\begin{equation}
\frac{d P^\alpha_{\rm em}}{du} = m_{\rm em} a^\alpha 
+ \frac{2}{3}\, q^2 a^2 u^\alpha,  
\label{8.4}
\end{equation}
where we have written $q^2/(2r) = m_{\rm em}$ in anticipation of our
eventual interpretation of the first term. Dirac now postulates that
the total momentum, $P^\alpha_{\rm em} + P^\alpha_{\rm mech}$, must be  
conserved: 
\begin{equation}
\frac{d P^\alpha_{\rm em}}{du} + \frac{d P^\alpha_{\rm mech}}{du} = 0.  
\label{8.5}
\end{equation}
Here, $P^\alpha_{\rm mech}$ is the mechanical momentum associated with
the point particle itself, and Eq.~(\ref{8.5}) becomes an equation of
motion if the correct relation between $P^\alpha_{\rm mech}$ and the
world-line quantities can be identified. This necessitates a second 
postulate. 

It would be tempting to make the identification $P^\alpha_{\rm mech} =
m_0 u^\alpha$, with $m_0$ describing the particle's material
mass. Combining this with Eqs.~(\ref{8.4}) and (\ref{8.5}), we would 
obtain $m a^\alpha = -\frac{2}{3} q^2 a^2 u^\alpha$, in which $m = m_0
+ m_{\rm em}$ would be interpreted as the particle's physical
mass. This, unfortunately, is a nonsensical result, because the
acceleration cannot be proportional to the four-velocity. 

We must therefore revisit our expression for the mechanical momentum,
and allow it to be a more complicated function of the world-line
quantities. We seek an expression of the form $P^\alpha_{\rm mech} = 
m_0 u^\alpha + c q^2 a^\alpha$, in which $c$ is a dimensionless
constant to be determined. (Other possibilities exist, but as Dirac
says, ``they are much more complicated than [this], so that one would
hardly expect them to apply to a simple thing like an electron''.)
Combining this with Eqs.~(\ref{8.4}) and (\ref{8.5}) yields $m
a^\alpha = -q^2 (c \dot{a}^\alpha + \frac{2}{3} a^2
u^\alpha)$. Demanding that the right-hand side be orthogonal to
$u^\alpha$ gives $c = -\frac{2}{3}$.  

Dirac's second postulate is therefore that the mechanical momentum
must have the form
\begin{equation}
P^\alpha_{\rm mech} = m_0 u^\alpha - \frac{2}{3}\, q^2 a^\alpha, 
\label{8.6}
\end{equation}
with $m_0$ representing the purely material contribution to the
particle's mass. This combines with the electromagnetic contribution
$m_{\rm em} = q^2/(2r)$ --- the electrostatic self-energy --- to form
the particle's physical mass $m$: 
\begin{equation}
m = m_0 + m_{\rm em} = m_0 + \frac{q^2}{2r}.
\label{8.7}
\end{equation}
Combining Eqs.~(\ref{8.4})--(\ref{8.7}) finally yields the
Lorentz-Dirac equation, 
\begin{equation}
m a^\alpha = \frac{2}{3}\, q^2 \bigl( \dot{a}^\alpha - a^2 
u^\alpha \bigr). 
\label{8.8}
\end{equation}
We have seen in Sec.~VI that the right-hand side of this equation is
produced by the half-retarded minus half-advanced potential, which is
regular on the world line. The role of the (singular) half-retarded
plus half-advanced potential is now clear: it serves to renormalize
the mass from $m_0$ to $m$; the difference is $m_{\rm em}$, the
electrostatic self-energy of the particle.      

It should be noted that while this derivation of the Lorentz-Dirac
equation stays quite close to spirit of Dirac's own derivation, it
differs from it in its technical aspects. Dirac employs a different
world-tube construction, based on spacelike geodesics (as opposed to
null geodesics) emanating from the world line. As a result, his
calculations are much more involved than the ones presented
here. Another consequence is that Dirac's expression for the
mechanical momentum differs from Eq.~(\ref{8.6}); it is the expected
$m_0 u^\alpha$. While Eq.~(\ref{8.6}) seems {\it ad hoc} and strange,
it is precisely the expression that results from a careful
consideration of the ``caps'' in Fig.~3 --- it is the mechanical
momentum that is appropriate for our particular world-tube
construction. The proof of this statement relies heavily on
distribution theory, and is presented in Ref.~\cite{TVW}. 

The final form of the Lorentz-Dirac equation is obtained by using  
the identity $a^2 = -\dot{a}_\alpha u^\alpha$ and inserting an
external-force term on the right-hand side of Eq.~(\ref{8.8}). This
gives 
\begin{equation}
ma^\alpha = F^\alpha_{\rm ext} + \frac{2}{3}\, q^2 
\bigl( \delta^\alpha_{\ \beta} + u^\alpha u_\beta \bigr) 
\dot{a}^\beta. 
\label{8.9}
\end{equation} 
If the external force is produced by an external electromagnetic field
$F^{\alpha\beta}_{\rm ext}$, then $F^\alpha_{\rm ext} = q
F^{\alpha\beta}_{\rm ext} u_\beta$. 

\section{Difficulties of the Lorentz-Dirac equation} 

A surprising feature of the Lorentz-Dirac equation is that it involves
the derivative of the acceleration vector. The equations of motion are 
therefore third-order differential equations for $z^\alpha(\tau)$, a
very unusual situation that necessitates some reflection. For example, 
an issue that must be addressed is the specification of initial data
for this third-order equation. In the usual case of second-order
differential equations, the initial data consists of the particle's
position and velocity at $\tau = 0$, and this information is
sufficient to provide a unique solution. A third-order equation
requires more, however, and it is not clear {\it a priori} what the
additional piece of initial data should be.  

To analyze this problem, let us consider the nonrelativistic limit of
the Lorentz-Dirac equation, which we write in the form
\begin{equation}
\bbox{a} - t_0 \dot{\bbox{a}} = \frac{1}{m}\, \bbox{F}_{\rm ext}, 
\label{9.1}
\end{equation}
where $\bbox{F}_{\rm ext}$ is an external force and
\begin{equation}
t_0 = \frac{2}{3}\, \frac{q^2}{m}
\label{9.2}
\end{equation}
is a constant with the dimension of time. For the purpose of this 
discussion we take $\bbox{F}_{\rm ext}$ to be a given function of 
time. It is easy to check that the general solution to Eq.~(\ref{9.1}) 
is 
\[
\bbox{a}(t) = e^{t/t_0} \biggl[ \bbox{b} - \frac{1}{m t_0}
\int_{-\infty}^t \bbox{F}_{\rm ext}(t') e^{-t'/t_0}\, dt' \biggr],
\]
where we assume that $\bbox{F}_{\rm ext}(t)$ goes to zero in the
infinite past, sufficiently rapidly that the integral is well
defined. The constant vector $\bbox{b}$ is not constrained {\it a
priori}; this is the third piece of data that must be specified to
completely determine the motion of the particle.  

To see how $\bbox{b}$ should be chosen, let us specialize to a
particularly simple case, in which the external force is turned on
abruptly at $t=0$ and stays constant thereafter: $\bbox{F}_{\rm
ext}(t) = \bbox{f} \theta(t)$, where $\bbox{f}$ is a constant
vector. In this case we have
\[
\bbox{a}(t) = e^{t/t_0} \biggl[ \bbox{b} - \frac{\bbox{f}}{m} \Bigl( 1
- e^{-t/t_0} \Bigr) \theta(t) \biggr], 
\]
and we see that for arbitrary choices of $\bbox{b}$, $\bbox{a} \sim
e^{t/t_0}$ for $t \gg t_0$. Even though the applied force is constant, 
the acceleration grows exponentially with time. This is the problem of
{\it runaway solutions}, which occurs also in the general case. We
notice, however, that these unphysical solutions can be eliminated if
we set $\bbox{b} = \bbox{f}/m$. Then we find  
\[
\bbox{a}(t) = \frac{\bbox{f}}{m} \Bigl[ \theta(-t) e^{t/t_0} +
\theta(t) \Bigr]. 
\]
Although $\bbox{a}$ is now sensibly behaved for $t > 0$, we see that
its behaviour is rather strange for $t < 0$: At a time $\sim t_0$ 
{\it prior} to the time at which the external force switches on, the  
acceleration begins to increase. This is the problem of 
{\it preacceleration}, which occurs also in the general case.  

\begin{problem}
Show that the only non-runaway solution to Eq.~(\ref{9.1}) is
\[
\bbox{a}(t) = \frac{1}{m}\, \int_0^\infty 
\bbox{F}_{\rm ext}(t + s t_0)\, e^{-s}\, ds.
\]
This solution displays preacceleration, because the acceleration at
a time $t$ depends on the external force acting at later times. 
\end{problem}  

The problems of runaways and preacceleration cast a serious doubt on 
the validity of the Lorentz-Dirac equation. The root of the problem
resides with the fact that we are trying to describe the motion of a
point particle within a purely classical theory of
electromagnetism. This cannot be done consistently. Indeed, a
point particle cannot be taken too literally in a classical context;
it must always be considered as an {\it approximation} to a
nonsingular, and extended, charge distribution. Essentially, the
difficulties of the Lorentz-Dirac equation come from a neglect to take 
this observation into account. 

Any extended charge distribution can be characterized by its total
charge $q$ and a number of higher multipole moments. For distances $r$ 
that are large compared with the body's averaged radius $\ell$, the
electromagnetic field is well approximated by the monopole term,
$q/r^2$, and in such circumstances, the point-particle description is
appropriate. For smaller distances, however, this description loses
its usefulness. The point-particle description is therefore an
approximation in which the internal structure of the charge
distribution can be considered to be irrelevant. This approximation is
limited to distances $r \gg \ell$, and the ``world line'' of a point
charge should properly be thought of as an extended world tube whose
dimensions are nevertheless smaller than any other relevant
scale. Such idealizations are common in physics; in hydrodynamics, for 
example, fluid elements are idealized as points.    

We should therefore examine the conditions under which Eq.~(\ref{9.1})
is compatible with the restrictions associated with a point-particle
description. Finite-size corrections can be incorporated into the
Lorentz-Dirac equation, which becomes   
\begin{equation}
\bbox{a} = \frac{1}{m}\, \bbox{F}_{\rm ext} + t_0 \dot{\bbox{a}}
+ O(t_0 \ddot{a} \ell). 
\label{9.3}
\end{equation}
This reveals that the original form of the equation will be accurate,
and the point-particle description valid, if $\ddot{a} \ell 
\ll \dot{a}$. To see what this means, let $a_c \sim F_{\rm ext}/m$ be
a characteristic value for the acceleration, and $t_c$ a
characteristic time scale over which the acceleration changes. Then
$\dot{a} \sim a_c/t_c$, $\ddot{a} \sim a_c/{t_c}^2$, and the
correction term will be small if 
\begin{equation}
\ell \ll t_c.
\label{9.4}
\end{equation}  
Thus, a necessary condition for the validity of the point-particle
description is that the time scale over which the acceleration changes 
must be much longer than the light-travel time across the charge
distribution.  

Another condition must be imposed. Any extended charge distribution
possesses an electrostatic self-energy given by $m_{\rm em} \sim
q^2/\ell$. This self-energy contributes to the total mass of the
charged body: $m = m_0 + m_{\rm em}$, where $m_0$ is the purely material
contribution to the mass. The classical theory will be consistent if
$m_{\rm em}$ does not exceed $m$: $m_0$ must be positive. Our second
condition is therefore $q^2/\ell \lesssim m$, or 
\begin{equation}
t_0 \lesssim \ell.
\label{9.5}
\end{equation}
Thus, the time constant $t_0$ must be smaller than the light-travel 
time across the charge distribution. 

Combining Eqs.~(\ref{9.4}) and (\ref{9.5}), we arrive at our most
important criterion: The point-particle description will be valid if 
and only if
\begin{equation}
t_0 \ll t_c.
\label{9.6}
\end{equation}
Thus, changes in the acceleration must occur over time scales that are 
long compared with $t_0$. This requirement is clearly violated by
runaway and preacceleration solutions. These predictions therefore
lie outside the domain of validity of Eq.~(\ref{9.1}). 

How do we then go about solving the runaway and preacceleration
problems? Before answering this, notice first that 
$\bbox{F}_{\rm ext}/m \sim a_c$ is the leading term on the right-hand
side of Eq.~(\ref{9.3}), that $t_0 \dot{\bbox{a}}$ is smaller by a
factor of order $(t_0/t_c)$, and that the remaining term is smaller
still, by a factor of order $(t_0/t_c)^2$. The Lorentz-Dirac equation
can therefore be written as 
\begin{equation}
\bbox{a} = \frac{1}{m}\, \bbox{F}_{\rm ext} + t_0 \dot{\bbox{a}} +
O({t_0}^2/{t_c}^2). 
\label{9.7}
\end{equation}
Our strategy will be to seek a replacement for this equation. (This 
strategy is the one adopted by Landau and Lifshitz; for a thorough
review of failed alternatives, see Ref.~\cite{FW}.) The new equation
will be equivalent to Eq.~(\ref{9.7}), in the sense that it will
differ from it only by terms of order $(t_0/t_c)^2$, but it will
be free of difficulties. In other words, the new equation will be
compatible with the restrictions associated with a point-particle
description. 

The replacement procedure is simple. We start with the leading-order
version of Eq.~(\ref{9.7}), 
\[
\bbox{a} = \frac{1}{m}\, \bbox{F}_{\rm ext} + O(t_0/t_c),
\]
and we differentiate. This gives
\[
\dot{\bbox{a}} = \frac{1}{m}\, \dot{\bbox{F}}_{\rm ext} +
O(t_0/{t_c}^2), 
\]
which we substitute back into Eq.~(\ref{9.7}). The result is
\[
\bbox{a} = \frac{1}{m}\, \bbox{F}_{\rm ext} + \frac{t_0}{m}\, 
\dot{\bbox{F}}_{\rm ext} + O({t_0}^2/{t_c}^2),
\]
which has the same degree of accuracy as Eq.~(\ref{9.7}). But since
this equation no longer involves $\dot{\bbox{a}}$, it does not give
rise to runaway and preacceleration solutions. 

Our conclusion is therefore that the modified Lorentz-Dirac equation,  
\begin{equation}
m \bbox{a} = \bbox{F}_{\rm ext} + t_0 \dot{\bbox{F}}_{\rm ext},  
\label{9.8}
\end{equation}
is much better suited to govern the motion of a point charge, within
the restrictions implied by the point-particle description. Equation
(\ref{9.8}) is formally equivalent to the original Lorentz-Dirac
equation, in the sense that they are both accurate up to terms of
order $(t_0/t_c)^2$. But because it is a second-order differential
equation for $\bbox{z}(t)$, Eq.~(\ref{9.8}) does not come with the
difficulties associated with the original equation. 

The method by which Eq.~(\ref{9.7}) was transformed into
Eq.~(\ref{9.8}) is an application of a general technique known as 
{\it reduction of order}. This technique can also be applied to the
relativistic equation, 
\[
a^\alpha = \frac{1}{m}\, F^\alpha_{\rm ext} + t_0 
\bigl( \delta^\alpha_{\ \beta} + u^\alpha u_\beta \bigr) 
\dot{a}^\beta. 
\]
To leading order, we have $a^\alpha = m^{-1} F^\alpha_{\rm ext}$, and
differentiation yields $\dot{a}^\alpha = m^{-1} 
F^\alpha_{{\rm ext}\, ,\beta} u^\beta$. Substituting this into the
original equation, we obtain  
\begin{equation}     
m a^\alpha = F^\alpha_{\rm ext} + t_0 
\bigl( \delta^\alpha_{\ \beta} + u^\alpha u_\beta \bigr) 
F^\beta_{{\rm ext}\, ,\gamma} u^\gamma, 
\label{9.9}
\end{equation}
the relativistic version of the modified Lorentz-Dirac equation. 

\begin{problem}
Show that if the external force is provided by an external
electromagnetic field $F_{\rm ext}^{\alpha\beta}$, then the 
modified Lorentz-Dirac equation takes the form
\begin{eqnarray*}
m a^\alpha &=& q F^\alpha_{{\rm ext}\, \beta} u^\beta 
+ q t_0 \biggl[ F^\alpha_{{\rm ext}\, \mu,\nu} u^\mu u^\nu 
\\ & & \mbox{}
+ \frac{q}{m} \bigl( 
\delta^\alpha_{\ \beta} + u^\alpha u_\beta \bigr) 
F^\beta_{{\rm ext}\, \mu} F^\mu_{{\rm ext}\, \nu} u^\nu \biggr].  
\end{eqnarray*}
\end{problem} 

\newpage

\section*{Acknowledgments}

I am grateful to Eanna Flanagan for his comments on an earlier draft 
of this manuscript. This work was supported by the Natural Sciences
and Engineering Research Council of Canada.

\end{document}